\documentclass[11pt,twoside]{article}
\usepackage{asp2004}\usepackage{psfig}\usepackage{epsf}
\usepackage{graphics}\usepackage{lscape}
\makeindex 
\begin{document}
\title{Frequency Evolution of the Gravitational Waves for Compact 
Binaries}
\author{B. Mik\'{o}czi}
\affil{Departments of Theoretical and Experimental Physics, 
University of Szeged, Szeged 6720, 
Hungary}

\begin{abstract} We present here a complete list of contributions 
to the gravitational wave 
frequency evolution from compact binaries on circular orbits up 
to the second post-Newtonian order 
by including the interaction of magnetic dipole moments, 
quadrupole-monopole interactions together 
with the spin-orbit, the spin-spin and the self-interaction spin 
contributions. We apply our 
results to the Manchester et al. model of the J0737-3039A-B 
double pulsar.
\end{abstract}

\section{Introduction} 

 Inspiralling compact binaries (like double-pulsar systems) emit 
gravitational radiation. These 
binary systems are the most promising source of the gravitational 
waves for the Earth-based 
interferometric detectors (LIGO, VIRGO, GEO, TAMA and AIGO). The 
coalescence of such compact 
binaries is preceded by a milder inspiral phase, for which the 
post-Newtonian approach provides a 
reliable description. This description is generally considered 
valid until the system reaches the 
innermost stable circular orbit (ISCO). The ISCO is not even 
precisely defined. For the special 
case of a non-rotating black hole the ISCO is r=6M and at r=M for 
a maximally rotating black hole 
(throughout the paper we use $G=c=1$). For neutron stars the 
gravitational wave frequency is 
800-1230 Hz \citep{Oechslin2004} at ISCO. The LIGO sensitivity 
is $~$1000 Hz.

The description of motion was given to 3.5 post-Newtonian (3.5 
PN) order accuracy, with the 
inclusion of spin-orbit effects (SO) \citep{Wex1995, RI1997, 
GPV1998} and their  first PN correction 
\citep{TOO2001}. Spin-spin (SS) 
\citep{Kidder1995, Gergely2000}, quadrupole-monopole (QM) 
\citep{Poisson1998, GK2003} 
and magnetic dipole-magnetic dipole (DD) \citep{IK2000, VKMG2003} 
and self-interaction spin term 
(SS-self) \citep{Gergely2000, MVG2005} contributions to 
the accelerations 
were also discussed. On the long 
run due to the emission of gravitational waves the orbit tends to 
circularize. Therefore we 
consider circular orbits, for which the gravitational wave 
frequency is twice the orbital 
frequency. We evaluate the rate of increase of $f$. This is given 
by the rate of change of the 
orbital angular frequency $\omega =\pi f$ under radiation 
reaction. We present the accumulated 
number of cycles left until the ISCO and the spin parameters in 
the Manchester et al. model of the 
J0737-3039A-B double pulsar.

\section{Frequency Evolution and Accumulated Number of Cycles}
 
In this section we summarize some of the results of 
\citep{MVG2005} on 
the frequency evolution and accumulated number of cycles for 
compact binaries on circular 
orbit. The radial projection of the acceleration defines the 
orbital angular velocity $\omega$. We 
expressed both the energy and the energy loss with $\omega$. Then 
the evolution of the radiative 
orbital angular frequency is:

\begin{eqnarray}
\left\langle \frac{d\omega }{dt}\right\rangle ^{circ} 
&=&\frac{96\eta
m^{5/3}\omega ^{11/3}}{5}\Biggl[1-\left( 
\frac{743}{336}+\frac{11}{4}\eta
\right) \left( m\omega \right) ^{2/3}+\left( 4\pi -\beta \right) 
m\omega  
\nonumber \\
&&+\Biggl(\frac{34103}{18144}+\frac{13661}{2016}\,\eta 
+\frac{59}{18}\,\eta
^{2}+\sigma \Biggr)\left( m\omega \right) ^{4/3}\Biggr],
\end{eqnarray}
where $\eta =m_{1}m_{2}/m^{2}$, $m=m_{1}+m_{2}$ and

\begin{equation}
\sigma =\sigma _{S_{1}S_{2}}+\sigma _{SS-self}+\sigma 
_{QM}+\sigma _{DD}.
\end{equation}
The quantities $\beta $, $\sigma _{S_{1}S_{2}}$, $\sigma 
_{SS-self}$, $
\sigma _{QM}$ and $\sigma _{DD}$ are the spin-orbit, spin-spin,
self-interaction spin, quadrupole-monopole and magnetic 
dipole-dipole
parameters, respectively: 

\begin{eqnarray}
\beta =\frac{1}{12}\sum_{i=1}^{2}\frac{S_{i}}{m_{i}^{2}}\left( 
113\frac{
m_{i}^{2}}{m^{2}}+75\eta \right) \cos \kappa _{i},  \nonumber \\
\sigma _{S_{1}S_{2}}=\frac{S_{1}S_{2}}{48\eta m^{4}}(-247\cos 
\gamma
+721\cos \kappa _{1}\cos \kappa _{2}),  \nonumber \\
\sigma _{SS-self}=\frac{1}{96m^{2}}\sum_{i=1}^{2}\left( 
\frac{S_{i}}{m_{i}}
\right) ^{2}\left( 6+\sin ^{2}\kappa _{i}\right) ,  \nonumber \\
\sigma _{QM}=-\frac{5}{2}\sum_{i=1}^{2}p_{i}\left( 3\cos 
^{2}\kappa
_{i}-1\right) ,  \nonumber \\
\sigma _{DD}=-\frac{5}{\eta m^{4}}d_{1}d_{2}\mathcal{A}_{0}.
\end{eqnarray}
The N, PN, SO, SS, 2PN and tail contributions were verified to 
agree with those given in \citep{PW1995}, the QM with those 
given in \citep{Poisson1998}, the DD 
with those given in \citep{IK2000} and the SS-self with those 
given in \citep{MVG2005}, 
respectively. The constant $\mathcal{A}_{0}$ is also given in 
\citep{VKMG2003}.

From here the accumulated number of gravitational wave cycles 
emerges as:
 
 \begin{eqnarray}
\mathcal{N} &=&\frac{1}{\pi \eta }\Biggl\{\tau ^{5/8}+\left( 
\frac{3715}{8064%
}+\frac{55}{96}\eta \right) \tau ^{3/8}+\frac{3}{4}\left( 
\frac{\beta }{4}%
-\pi \right) \tau ^{1/4}  \nonumber \\
&&+\Biggl(\frac{9275495}{14450688}+\frac{284875}{258048}\eta 
+\frac{1855}{%
2048}\eta ^{2}-\frac{15\sigma }{64}\Biggr)\ \tau ^{1/8}\Biggr\}.
\end{eqnarray}
where the dimensionless time variable $\tau =\eta (t_{c}-t)/5m$ 
is related
to the time $(t_{c}-t)$ left until the final coalescence. The 
tail and 2PN
contributions agree with those given in \citep{BFIJ2002}.

\begin{table}[!ht]
\caption{Accumulated number of gravitational wave cycles.}
\smallskip
\begin{center}
{\small
\begin{tabular}{llll}
\tableline
\noalign{\smallskip}
PN Order & $J0737-3039$ & $B1913+16$
& $BH-BH$ \\
& $1.337M_{\odot }$
& $1.387M_{\odot }$
& $10^{4}M_{\odot }$ \\
& $1.250M_{\odot }$
& $1.441M_{\odot }$
& $10^{5}M_{\odot }$ \\
\noalign{\smallskip}
\tableline
\noalign{\smallskip}
$f_{in}(Hz)$ & $10$ & $10$
& $4.199\times 10^{-4}$ \\ 
$f_{fin}(Hz)$ & $1000$ & $1000$ & $3.997\times 10^{-2}$ \\ 
\noalign{\smallskip} 
\tableline
\noalign{\smallskip}$N$ & $18310$ & $15772.1$ & $21058$ \\ 
$PN$ & $475.8$ & $435$ & $677$ \\ 
$SO$ & $17.5\beta $ & $16.5\beta $ & $36\beta $ \\ 
$SS-self,SS,QM,DD$ & $-2.1\sigma $ & $-2.1\sigma $ & $-5\sigma $ 
\\ 
$Tail$ & $-208$ & $-206$ & $-450$ \\ 
$2PN$ & $9.8$ & $9.5$ & $18$ \\
\noalign{\smallskip}
\tableline
\end{tabular}
}
\end{center}
\end{table}

\begin{table}[!ht]
\caption{The spin parameters for the two solutions (JR1 \& JR2) 
of the Jenet-Ransom model
and for two solutions (M1 and M2) of the model of the Manchester 
et al. representing the
binary pulsar J0737-3039.}
\smallskip
\begin{center}
{\small
\begin{tabular}{ccccc}
\tableline
\noalign{\smallskip}
Spin parameters & JR1 & JR2
& M1 & M2\\
& [$\kappa _{1}=167^{\circ }$]
& [$ \kappa _{1}=90^{\circ }$]
& [$ \kappa _{1}=14^{\circ }$]
& [$ \kappa _{1}=60^{\circ }$] \\
\noalign{\smallskip}
\tableline
\noalign{\smallskip}
$\beta \;$ & $-0.166$ & $0.001$ & $0.168$ & $0.087$ \\
$\sigma _{S_{1}S_{2}}\;(10^{-4})$ & $-0.372$ & $0$ & $0.371$& 
$0.191$ \\
$\sigma _{SS-self}$\/$\;(10^{-4})$ & $0.298$ & $0.345$ & $0.298$ 
& $0.333$ \\
\noalign{\smallskip}
\tableline
\end{tabular}
}
\end{center}
\end{table}

In Table 1 we present numerical results for two well-known binary 
neutron star systems, the double 
pulsar J0737-3039 \citep{Burgay03, Lyne2004}, 
Hulse-Taylor pulsar B1913+16 and one 
example of galactic black hole-galactic 
black hole binary.
 
\section{Application for the J0737-3039A-B Double Pulsar}

The double-pulsar J0737-3039 has become an important 
astrophysical laboratory for testing 
gravitational physics. The measurements on this system where done 
by the Parkes 64-m 
radio-telescope at three frequencies: 680, 1390 and 3030 MHz 
\citep{Manchester2004}.

We evaluate the spin parameters $\sigma _{S_{1}S_{2}}$ and 
$\sigma _{SS-self}$ for the case of
the double pulsar J0737-3039. The neutron stars in this double 
pulsar have average radii $15$ km, 
pulse periods of $22.7$ ms and $2773.5$ ms. Tha $\kappa _{1}$ 
angle is between the spin axis and 
angular momentum. The Jenet-Ransom model \citep{JR2004} 
has two possible values: $\kappa _{1}=167^{\circ}\pm 
10^{\circ }$ (JR1), and $\kappa _{1}=90^{\circ }\pm 10^{\circ }$ 
for pulsar A (JR2).\ The other 
angle $\kappa _{2}$ for pulsar B in principle can be determined 
by solving numerically the spin 
precession equations. According to \citep{Kaspi2004} it is 
likely that wind-torques from the 
energetically dominant component have driven the spin axis of the 
other
component to align with the direction of $\mathbf{L}$, causing 
$\kappa
_{2}=0 $. Therefore $\gamma =\kappa _{1}$. We give the estimates 
for the spin parameters in Table 
2.

Besides the JR1 and the JR2 solutions we present here for the 
first time an model \citep{Manchester2004}. According to this 
model the measurements the favorize 
maximally allowed the angle 
$\kappa _{1}=60^{\circ}$. The best-fit solution (M1) is located 
at $\kappa _{1}=14^{\circ}$. We 
also present results for the maximally allowed value of $\kappa 
_{1}$ (M2). As it can  be seen in 
this model the proper spin-spin contribution is comparable with 
the self-interaction spin 
contribution.

\section{Conclusions}

We have presented the complete set of contributions up to second 
post-Newtonian order 
(PN, SO, SS, QM, DD, tail and 2PN) to the evolution of 
gravitational wave frequency and to the accumulated number
of gravitational wave cycles left until the final coalescence, 
with the
inclusion the self-interaction spin terms (SS-self).

We have shown that on the Manchester et al. model of the 
J0737-30309A-B double pulsar the 
self-interaction spin contributions are comparable with spin-spin 
contributions.
  
This work was supported by OTKA no. TS044665 and T046939 grants.

{}


\begin{thebibliography}{}

\bibitem[Blanchet et al. 2002]{BFIJ2002} Blanchet, 
L., Faye G., Iyer, B. R., Joguet, B. 2002, Phys. Rev. D65, 
061501
\bibitem[Burgay et al. 2003]{Burgay03} Burgay, M., et al. 2003, 
Nature 426, 531
\bibitem[Gergely 2000]{Gergely2000} Gergely, L. \'{A}. 2000, 
Phys. Rev. D62 024007; ibid. D61, 024035
\bibitem[Gergely \& Keresztes 2003]{GK2003} Gergely, L. \& 
\'{A}., Keresztes, Z. 2003, Phys. Rev. D67, 024020
\bibitem[Gergely et al. 1998]{GPV1998} Gergely, 
L. \'{A}.,  Perj\'{e}s, Z. I., Vas\'{u}th, M. 1998 Phys. Rev. 
D57, 876; ibid. D57, 3423; ibid. D58, 124001
\bibitem[Ioka \& Taniguchi 2000]{IK2000} Ioka, K. \& Taniguchi, 
T. 2000, Phys. Rev. 537, 327
\bibitem[Jenet \& Ransom 2004]{JR2004} Jenet, F. A. \& Ransom, S. 
M. 2004, Nature 428, 919
\bibitem[Kaspi et al. 2004]{Kaspi2004} Kaspi, V.M., et al. 2004, 
ApJ. 613, 137
\bibitem[Kidder 1995]{Kidder1995} Kidder, L. 1995, Phys. Rev. 
D52, 821
\bibitem[Lyne et al. 2004]{Lyne2004} Lyne, A. G. et al. 2004, 
Science 303, 1153
\bibitem[Manchester et al. 2004]{Manchester2004} Manchester, R. 
N., et al. 2005, ApJ. 621 L49
\bibitem[Mik\'{o}czi et al. 2005]{MVG2005} 
Mik\'{o}czi, B., Vas\'{u}th, M., Gergely, L. \'{A}. 2005, Phys. 
Rev. D71 124043
\bibitem[Oechslin et al. 2004]{Oechslin2004} Oechslin, R., et al. 
2004, Mon. Not. Roy. Astron Soc. 349, 1469
\bibitem[Poisson 1998]{Poisson1998} Poisson, E. 1998, Phys. Rev. 
D57, 5287
\bibitem[Poisson \& Will 1995]{PW1995} Poisson, E., Will, C. M. 
1995, Phys. Rev. D52, 848
\bibitem[Rieth \& Schafer 1997]{RI1997} Rieth, R. \& Schafer, G. 
1997, Class. Quantum Grav, 14, 2357
\bibitem[Tagoshi et al. 2001]{TOO2001} Tagoshi, H., 
Ohashi, A., Owen, B. J. 2001, Phys. Rev. D63 044006 
\bibitem[Vas\'{u}th et al. 2003]{VKMG2003} Vas\'{u}th, M., 
Keresztes, Z., Mih\'{a}ly, A., 
Gergely, L. \'{A}. 2003, Phys. 
Rev. D68, 124006
\bibitem[Wex 1995]{Wex1995}Wex, N. 1995, Class. Quantum Grav. 12, 
983
\end{thebibliography}
\end{document}